%% file: bare_jrnl.tex
\documentclass[journal]{IEEEtran}
\usepackage{xcolor}
\usepackage{amsmath}
\usepackage{algorithmic}
\usepackage{url}
\usepackage[most]{tcolorbox}
\usepackage{hyperref}
\usepackage{csquotes}
\usepackage{booktabs}

\begin{document}
%
\title{Unmasking the Shadows: Pinpoint the Implementations of Anti-Dynamic Analysis Techniques in Malware Using LLM}
%
%
%

\author{
    \IEEEauthorblockN{Haizhou Wang\IEEEauthorrefmark{1}, Nanqing Luo\IEEEauthorrefmark{1}, Xusheng Li\IEEEauthorrefmark{2}, Peng Liu\IEEEauthorrefmark{1}}\\
    \IEEEauthorblockA{\IEEEauthorrefmark{1}Pennsylvania State University
    \\\{hjw5074, nkl5280, pxl20\}@psu.edu}
    \\\IEEEauthorblockA{\IEEEauthorrefmark{2}Vector 35 Inc.
    \\xusheng@vector35.com}
    
}

\maketitle

\input{sections/abstract}


%
\IEEEpeerreviewmaketitle

\input{sections/introduction}

\input{sections/background}

\input{sections/problem}

\input{sections/method}

\input{sections/evaluation}

\input{sections/discussion}
\input{sections/relatedwork}

\input{sections/conclusion}


%

\appendices
\input{sections/appendix}





\ifCLASSOPTIONcaptionsoff
  \newpage
\fi



\bibliographystyle{IEEEtran}
\bibliography{ref}

%








\end{document}

%% file: sections/abstract.tex
\begin{abstract}
Sandboxes and other dynamic analysis processes are prevalent in malware detection systems nowadays to enhance the capability of detecting 0-day malware.
Therefore, techniques of anti-dynamic analysis (TADA) are prevalent in modern malware samples, and sandboxes can suffer from false negatives and analysis failures when analyzing the samples with TADAs.
In such cases, human reverse engineers will get involved in conducting dynamic analysis manually (i.e., debugging, patching), which in turn also gets obstructed by TADAs.
In this work, we propose a Large Language Model (LLM) based workflow that can pinpoint the location of the TADA implementation in the code, to help reverse engineers place breakpoints used in debugging.
Our evaluation shows that we successfully identified the locations of 87.80\% known TADA implementations adopted from public repositories.
In addition, we successfully pinpoint the locations of TADAs in 4 well-known malware samples that are documented in online malware analysis blogs.

\end{abstract}

%% file: sections/introduction.tex
\section{Introduction}



Malware detection in real-world is achieved through a complex human-in-the-loop system with reverse engineers playing an essential role.
In terms of quantity, the majority of the malware samples are detected and blocked by file signatures, including inline firewall blocking and anti-virus software.
This is because most samples in the wild are 1-day or n-day samples, so their file signatures are already documented, either automatically or manually.
Obviously, any signature-based methods are subject to 0-day malware attacks, and therefore different methods need to be adopted to detect and block 0-day malware.
Although it is possible to detect 0-day malware through advanced static analysis, in practice this strategy can suffer from non-scalability and false negatives.
For example, some samples are only at the initial stage of infection and will do no harm unless subsequent payloads are downloaded, which can only be confirmed to be malicious by concretely executing the samples.
Therefore, many industry-level malware detection systems rely on automated dynamic analysis, or sandboxes~\cite{kirat2014barecloud, kruegel2016build}, to detect 0-day malware using behavior signatures and rules.
Unfortunately, most malware authors are fully aware of the existence of sandboxes and dynamic analysis efforts, and therefore many of them nowadays will implement Techniques of Anti-Dynamic-Analysis (TADA) in their malware to evade dynamic analysis-based detection.

\begin{figure}[ht]
    \centering
    \includegraphics[width=0.83\linewidth]{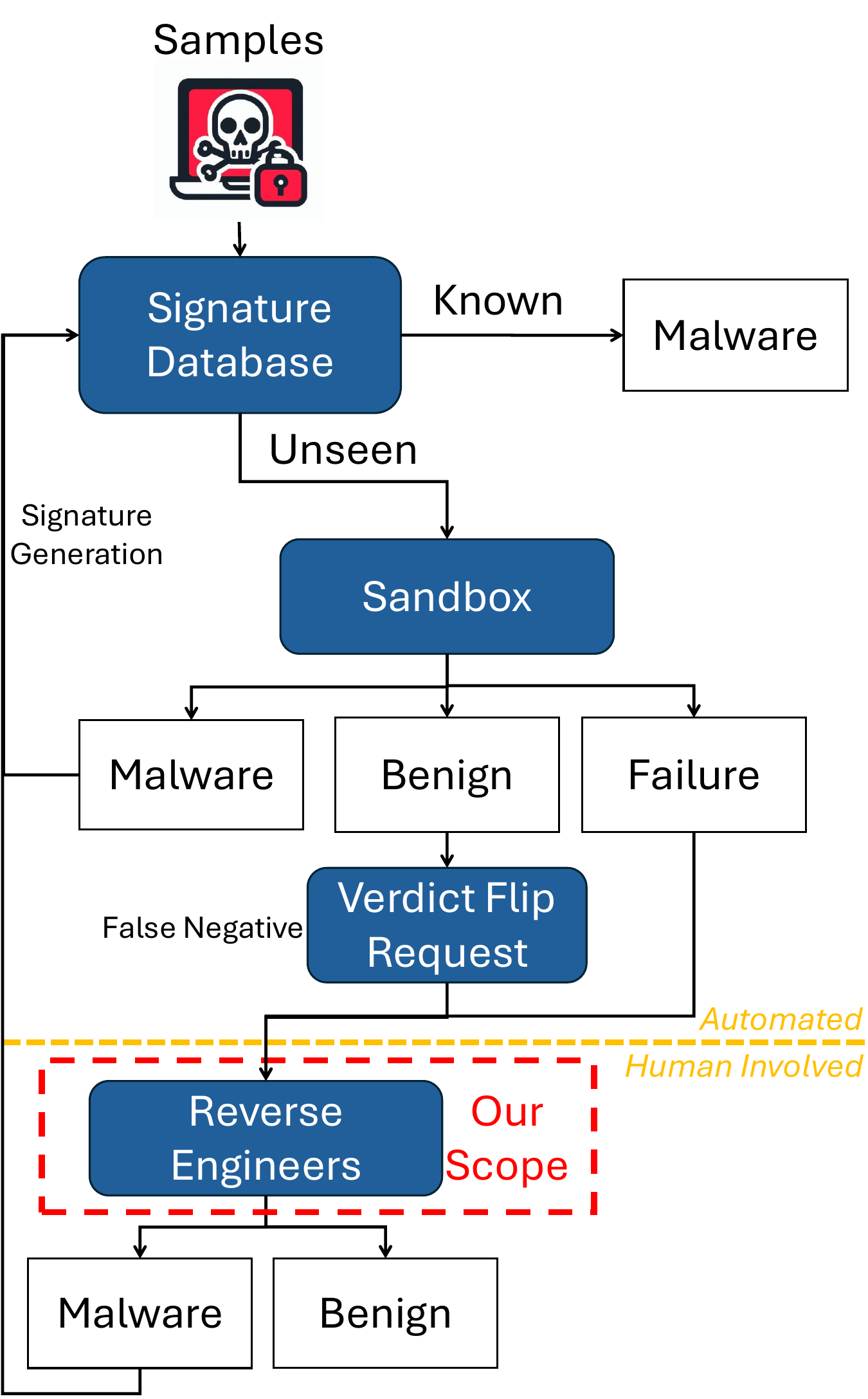}
    \caption{Practical Malware Detection System}
    \label{fig:malware_detection_system}
\end{figure}

As we will elaborate in \autoref{sec:background:tada}, malware authors usually have \textbf{two goals} with TADAs implemented: evading the dynamic analysis-based detection and impeding reverse engineering effort, respectively.
These two goals are directly motivated by the process of detecting malware samples in the real world, which is illustrated in
\autoref{fig:malware_detection_system}.
The top half of the figure is fully automated, and the bottom half involves human labor.
When a sample arrives at the system, the first step is usually to check the existing signature and hash databases to see if the sample is already known to be malicious.
If the signature or hash has a match, then the system can immediately conclude that the sample is malicious.
Despite the disadvantage of the file signature-based detection (i.e. unable to handle 0-day), it is still worthwhile to have it in the detection system to block 1-day and n-day malware, because it is extremely fast and affordable.
To keep the databases updated, whenever a malware verdict is generated in the later part of the system (meaning that the sample has not been seen), it will be used to update the databases.
Subsequently, in cases where there is no match for the given sample, it will then be transferred to a sandbox, where the sample will be dynamically analyzed.
The sample will be executed and monitored without human intervention.
The sandbox analysis usually has 3 outcomes: malware verdict, benign verdict, and analysis failure.
Since detection is achieved through conservative behavior heuristics and rules, the malware verdict is usually true positive.
In contrast, a benign verdict can be false negative, due to either TADAs or missing OS components that cause the sample not to behave maliciously in the sandbox environment.
These false negatives are usually found later, and the system will receive a verdict flip request.
As for analysis failures, they are the cases where the sample failed to launch at all or crashed during the execution.
The most common reason for false negative and analysis failures is the presence of TADA, as sandboxes are usually virtual machines (VM) that conduct automatic dynamic analysis.
Consequently, at the point of failures or verdict flip requests, human intervention is necessary, and the sample is then sent to a reverse engineer for the final verdict.

Back to the discussion of the goals of TADAs, the ultimate goal of the malware authors is to delay the detection of their malware as much as possible, so that more victims could be infected.
Based on \autoref{fig:malware_detection_system}, it is obvious that the best situation for malware authors' interest is when a malware sample bypasses the signature database, bypasses sandboxes, and eventually reaches the hands of the reverse engineer, being analyzed for an extended period of time.
Consequently, the first goal of TADAs is to bypass the sandbox detection, which is usually achieved by detecting the artifacts (e.g. analysis programs, common VM hardware names) in the sandbox.
To fight back, there are numerous previous works studying how to prevent malware from bypassing the sandboxes in the first place~\cite{d2020dissection,kruegel2016build,kirat2014barecloud,kruegel2014full,khasawneh2017rhmd,lindorfer2011detecting}, most of which can be summarized in one single word: transparency.
The idea is simple: as long as the sandbox is transparent (i.e., the sandbox is not discernible with the real machine), the malware will trigger malicious behaviors.
As we will elaborate in \autoref{sec:problem:motivation}, despite the theoretical feasibility of complete transparency, in the real world pursuing transparency is essentially a trade-off between cost and transparency enhancement.
Therefore, it is very common that a malware sample eventually will get to the hands of a reverse engineer, which corresponds to the second goal of TADAs:
impeding the reverse engineering efforts to delay the analysis.
Thus, from the reverse engineers' perspective, the faster they can find the TADA implementation (in the code), the better so that they can bypass or patch the TADAs during their dynamic analysis.
Although there are existing works studying detecting malware with TADA~\cite{lindorfer2011detecting,aboaoja2023dynamic,kirat2015malgene}, most of them focus on the detection of anti-analysis and malicious behaviors rather than the problem of identifying the locations of TADAs.

As shown in \autoref{fig:malware_detection_system}, the scope of this research is not to detect the malware with TADAs (i.e., evasive malware) in general; rather we focus on complementing the current reverse engineering efforts on evasive malware, reducing the time and labor efforts of reverse engineers.
Specifically, this research aims to identify the locations of the TADA in the code for breakpoints used in manual dynamic analysis (i.e. debugging) during the reverse engineering efforts.
However, pinpointing the locations of TADA implementations can be challenging.
First, there are a variety of TADAs: checking hardware (e.g., CPU cores, memory size, PCIe devices, power capabilities, fans, etc.), checking running processes, checking filesystems, checking user traces, etc.
Second, not only can malware adopt different techniques, but also can they use different implementations. Taking the number of CPU core as an example, it can be checked via excessive number of implementations (e.g., CPUID, WMI Query, different APIs, etc.), all up to malware authors' creativity.
This diversity of the TADA implementations essentially makes the detection very prone to false negatives.

In this paper, we present a workflow that leverages advanced static analysis and a Large Language Model (LLM) to pinpoint the locations of TADA in the code, given a malware binary executable.
The workflow is based on two key insights.
First, an LLM can be leveraged to address a primary limitation of current rule-based static analysis for TADAs.
That is, the rule-based approaches are, in principle, not scalable when new TADAs and TADA implementations keep on emerging.
Second, the semantic gap between disassembled code and natural languages results in unsatisfactory responses when the LLM is directly asked to examine disassembly code, which could be bridged by advanced static analysis.
Our major contributions are as follows:
\begin{itemize}
    \item Propose a LLM-based workflow that can pinpoint the locations of the TADA in a binary executable automatically.
    \item Construct the useful features that can enable the LLM to do the detection.
    \item Evaluate our method using real-world malware samples of popular families.
\end{itemize}
Specifically, we successfully identify the locations of 87.80\% of known TADA implementations obtained from public repositories.
In addition, we successfully pinpoint the locations of TADAs in 4 well-known malware samples that are documented in online malware analysis blogs.

%% file: sections/background.tex
\section{Background} \label{sec:background}

\subsection{Techniques of Anti Dynamic Analysis (TADA)} \label{sec:background:tada}
Existing works~\cite{afianian2019malware,jadhav2016evolution,gao2014survey} had surveyed TADA comprehensively.
From an attacker's perspective, the reason for implementing TADA in their malware is two-fold: evade dynamic analysis based detection and analysis obstruction.
Evading dynamic analysis based detection is to ensure the malware will not be blocked so that the malicious goals can be achieved; whereas analysis obstruction is the effort to delay the reverse engineer's understanding the logic of the malware as well as understand the root cause of analysis evasion, if any.

Dynamic analysis usually happens in sandboxes based on Virtual Machines (VM) and/or other monitored environments.
Since sandbox is automated dynamic analysis, it is omnipresent in malware detection systems.
To evade dynamic detection, malware must not behave maliciously in sandboxes, rendering the necessity of detecting the sandbox environment.
\cite{oyama2018trends} shows the trend that fingerprinting and artifact detection is becoming popular in sandbox evasion, not only because it is highly effective, but also because the implementation could be extremely diverse.
As a counter-measure, sandbox developers usually will try to hide the artifacts indicating sandbox and VM environment~\cite{kirat2014barecloud, kruegel2016build,d2020dissection,kruegel2014full,khasawneh2017rhmd}.
With this being the current status, essentially, a good sandbox is usually transparent, but the transparency is not free: the more transparent, the more costly.
For example, cost-effective sandboxes are usually VM-based, because VMs are easy to create and destroy, no requirement of isolated hardware, and easy to scale.
However, VM-based sandboxes are relatively less transparent, especially when necessary monitoring tools need to be integrated into the guest OS.
In contrast, the most transparent sandboxes is bare-metal hardware based, but that would cost much more than a VM-based sandbox.
Thus, due to this real-world cost effectiveness issue, for malware authors, implementing TADAs to evade detection is still considered to be worthwhile.

Another reason of malware having TADA is analysis obstruction.
In short, failed analysis and false negative could happen when detecting 0-day malware dynamically, either due to the TADAs implemented by the malware author or missing runtime requirements.
Therefore, eventually manual analysis by reverse engineer becomes necessary to confirm the verdict of a sample for which automatic analysis failed.
Reverse engineers usually rely on the debuggers and other analysis software such as \texttt{Process Monitor} to understand the behavior of a sample.
Accordingly, many TADA implemented is targeting to detect the debuggers and analysis programs (i.e., anti-debugging)~\cite{chen2008towards}.

\subsection{Categorization of TADAs} \label{sec:background:category}
Depending on the context, analysts tend to categorize TADAs differently.
There are two major ways to categorize TADAs: by tactic or by implementation.
For sandbox builders, they often categorize TADAs by their tactics, because sandbox builders usually focus on how to hide the sandbox artifacts and make their analysis tools more transparent.
For example, in \cite{afianian2019malware}, authors categorize the TADAs based on the evasion goals.
Whereas for reverse engineers, they tend to categorize TADAs by their implementations, because reverse engineers usually need to identify the locations of the code implementing TADAs.

\textbf{Categorization based on tactics.} The most popular strategy to categorize TADAs is based on their \textit{tactics}.
In \texttt{MITRE ATT\&CK} tactics, there are two sub-tactics in Defense Evasion that is related to TADA:
\begin{itemize}
    \item T1622: Debugger Evasion
    \item T1497: Virtualization/Sandbox Evasion
\end{itemize}
Both T1622 and T1497 aim to detect dynamic analysis efforts, with slight different focuses.
On one hand, T1622 is debugger evasion, which is usually aiming to detect and impede human analysts reverse engineering efforts (debuggers are usually used by a human); on the other hand, T1497 is more about impeding automated dynamic analysis for malware evasion and detection.
For example, the most well-known debugger detection on Windows platform is to check the Process Environment Block (PEB), where there exists multiple indicators that may imply the presence of a debugger and tracer.
In general, we can categorize TADAs into 4 categories based on their tactics:
\textbf{debugger evasion}, \textbf{sandbox evasion}, \textbf{VM evasion}, and \textbf{analysis tool evasion}.
Regarding the differences between VM and sandbox evasion, a VM does not have to be a sandbox (e.g. VMs used by reverse engineers), and a sandbox does not have to be a VM (e.g. bare-metal sandboxes, emulator sandboxes).

\textbf{Categorization based on implementations.} A less popular way to categorize TADA is based on the \textit{implementations}.
Since the implementations of TADAs are not necessarily correlated to their tactics, the implementations of TADA with different tactics can be extremely alike.
For instance, by altering very little can an implementation of a running debugger detection technique turn into a VM guest addition detection.
If the running debugger is detected by finding a well-known debugger process name,
all the malware writer needs to do to detect VM guest addition is to switch the name of the process to detect.
From a program analysis point of view, the code involving detecting running debugger and detecting running virtual machine guest addition could be virtually identical, but the purpose is very different, which is purely based on the name of the process being detected.
By implementation, TADAs can be categorized into: \textbf{assembly based}, \textbf{direct API based}, and \textbf{indirect API based}.
As the name suggested, assembly-based implementations usually achieve the TADA goals using several instructions.
Usually, this kind of TADA is implemented by adding inline assembly code.
Direct API-based TADAs leverage the APIs that can directly inform the program if it is executing under an analysis environment, such as the famous \texttt{IsDebuggerPresent} API.
Usually direct API-based TADA will not involve many API calls: one or a few will be sufficient for the conclusion.
In contrast, indirect API-based TADA uses APIs that seem to not be directly related to the purpose of anti-dynamic analysis (e.g. check the current username).
Usually indirect API-based TADA implementation will involve a sequence of API calls, followed by logic checking their return values.

\subsection{LLM For Binary Analysis} \label{sec:background:llm}
LLMs we refer to today are usually Generative Pre-trained Transformers (GPT)~\cite{gpt3}, which take tokenized natural language inputs and generate language output.
These LLMs have been showing impressive performance in source code analysis, as technically programming language is just a more structured natural language that will only apply to a limited scope.
However, when applying LLMs for binary analysis, it is necessary to conduct an extensive amount of feature engineering, because raw binary code are very different from the natural language.
Even disassembled binary code, or assembly language, is significantly different.

There are few existing works that have investigated the potential of LLMs assisting binary analysis~\cite{liu2023harnessing, gao2023far, rong2024disassembling, ye2024detecting}.
Most of them adopted decompilation of the binary code and used the pseudo C code as the starting point to generate inputs for LLMs.
Such a strategy is very effective but could still be vulnerable to decompiling mistakes, such as incorrect type inference and broken data flow.
Therefore, it is a good tactic to rely on the decompiled code as little as possible.

%% file: sections/problem.tex
\begin{figure*}
    \centering
    \includegraphics[width=0.88\linewidth]{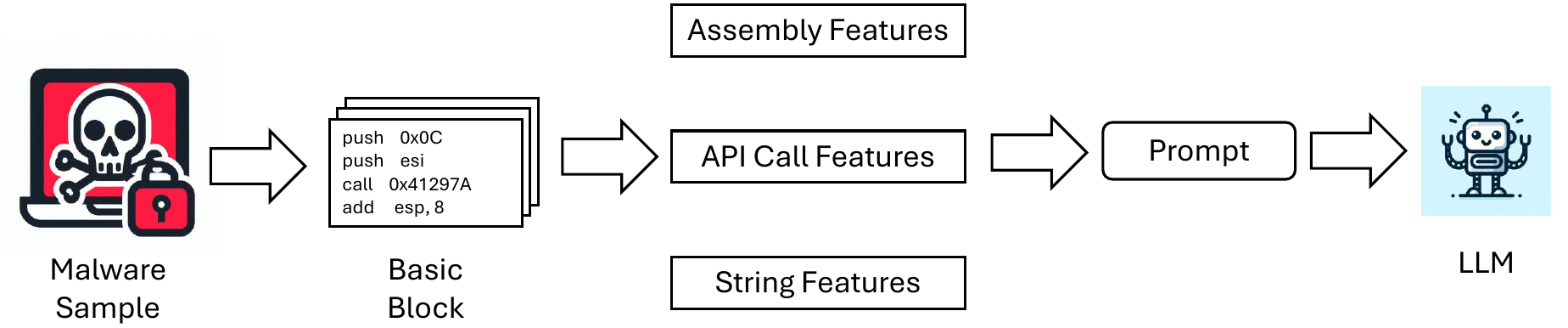}
    \caption{Overview of our workflow}
    \label{fig:workflow}
\end{figure*}

\section{Motivation and Problem Statement} \label{sec:problem}
\subsection{Motivation}\label{sec:problem:motivation}

Due to the diversity of TADAs and cost-effectiveness concerns when designing the sandboxes, it is still common for malware to evade sandbox detection and eventually in need of manual reverse engineering.
Common reverse engineering process for binary executables involves two major steps: 1) advanced static analysis for interesting breakpoints and memory locations, and 2) dynamic analysis using a debugger for behavior study (using breakpoints found previously).
Dynamic analysis using debugger is usually necessary throughout the reverse engineering efforts, either to search for malicious behaviors or confirm conclusions drawn from advanced static analysis.
As described in \autoref{sec:background:tada}, malware authors implement TADAs not only to evade detection but also to impede and delay human reverse engineering analysis.
Therefore, it is extremely likely that when a reverse engineer investigates an evasive sample, he/she will encounter TADAs stopping them from properly analyzing the sample.

This work is motivated by the fact that the real-world malware detection system shown in \autoref{fig:malware_detection_system} still requires a significant amount of human effort that can be tedious for the reverse engineer, yet most parts of the system are already automated (top side in \autoref{fig:malware_detection_system}).
Based on the common two-step reverse engineering process, we aim to propose a workflow that can suggest a set of breakpoints near the implementation of TADA during the advanced static analysis stage, so that the analysis can apply patches whenever these breakpoints are triggered during the dynamic analysis stage.

\subsection{Advantages of LLMs}
\label{sec:problem:llm}
Introduced in \autoref{sec:background:category}, the implementations of the TADAs can be categorized into 3 kinds.
As we will show later in \autoref{sec:evaluation}, the majority of the TADAs are implemented through indirect APIs, against which are extremely tricky to defend, because never can human analysts predict what aspects of the runtime environment malware authors are checking.
Based on various analysis reports~\cite{gao2014survey, vast_ocean, jadhav2016evolution}, malware may inspect username, folder and file names, disk volume number, recent file used, number of CPU cores, current running process names, current running process command lines, etc.
Since modern computers and OSes are extremely complicated, the list of sandbox artifacts is endless, and consequently, it is not possible to create 100\% transparent sandboxes.

If we scrutinize the implementations of TADAs carefully, it is notable that the majority of the runtime checks involve strings.
For example, inspecting file names, user names, WMI (Windows Management Instrumentation) queries, and all other similar artifacts will presumably involve strings.
Due to the volatile nature of strings, detection of TADA could suffer from false negatives when using rule-based methods, which may be tackled by adopting LLMs.
Despite the controversy about LLMs' ability to truly understand humans, LLMs do show a promising ability to understand human languages, which makes them a perfect candidate for digesting the strings used in TADA implementations.

\subsection{Problem Statement} \label{sec:problem:problem}
As motivated in \autoref{sec:problem:motivation}, our goal is to suggest a set of breakpoints "near" the implementation of TADA, so that the reverse engineer can bypass the TADAs and conduct root cause analysis on sandbox failures.
Now we have our problem statement as follows:
\begin{displayquote}
\textit{Given a malware sample with TADA, how to reduce the time and labor efforts of the reverse engineer by providing a set of breakpoints.}
\end{displayquote}

In this paper, we decide to perform our analysis at the basic block (BB) level, and correspondingly, the concept of "near" will be a small number of BB away from the BB where the breakpoint is located, on the control flow graph (CFG).
Specifically, we want to place the breakpoints at the BBs that are parts of the TADA implementations.


%% file: sections/method.tex
\section{Method} \label{sec:method}

\subsection{Overview} 
\label{sec:method:overview}
To identify the breakpoints for human analysts, they need to be located near the implementation of the TADAs.
Therefore, our goal can also be viewed as detecting the location of the TADA.

\noindent \textbf{What is the granularity of the analysis?}
The breakpoints are usually placed at the beginning of instructions so that the granularity of the breakpoints is at the instruction level.
However, for the sake of detecting TADA implementations, instruction level could be improper, as the amount of information carried in an single instruction could be too little.
The most obvious example are API calls, which usually involves more than one instructions if any arguments are presented.
On the other hand, function-level could be too large, as placing a breakpoint at a function could be futile, in worse cases requiring human analyst manually inspecting thousands of instructions. 
Therefore, to include sufficient information for detection while keeping the definition of the breakpoint location reasonably precise, our detection granularity is at the basic block (BB) level.

\noindent \textbf{What input fits LLM well?}
This paper focuses on reverse engineering of binary programs, which is clearly inappropriate to be directly integrated into the input prompt.
Essentially, both open-source and proprietary LLMs are trained using natural languages in addition to some program languages so that the LLMs cannot handle reverse engineering tasks from scratch.
Common strategies to address this type of challenge include few-shot learning and fine-tuning, but given the diversity of the TADA implementations, such strategies face the challenge of collecting high-quality training datasets.
Therefore, to ensure the LLM is well exploited, instead of providing hex strings of raw binary or assembly code, we first create necessary features to find TADA implementations for each BB, that are described in the form of natural language.
Despite our BB level granularity, it does not mean that our analysis is at BB level only, while we leverage control flow and data flow analysis to extract the features we need.

\noindent \textbf{Workflow summary.}
At a higher level, we first conduct program analysis to extract the features that are necessary to determine whether a BB is TADA-related or not.
The necessity and the selection of the features are based on domain knowledge, which will be explained in detail in the following subsections.
Those features will be then used to build prompts and subsequently sent to LLM for the final output.
Although the output of our workflow is a binary decision for each BB, we prompt the LLM so that a response of ratings (of how related the BB is to TADA implementations) is generated for each BB.

Our workflow only involves static analysis, which is driven by the common reverse engineering practices mentioned at the beginning of \autoref{sec:problem:motivation}.
Better, static analysis is immune to the coverage issue caused by having multiple TADAs implemented in one sample: the malware sample will exit or sleep after the first check fails, not executing other TADAs it may have.
If we had adopted dynamic analysis, we would have had to patch the binary after detecting one TADA before detecting another.

The major steps are illustrated in \autoref{fig:workflow}.
Starting from a malware sample, we first construct the control flow graph (CFG) and extract all the BBs from the executable.
BBs consist of instructions, from which along with the static analysis over the whole binary we extract three kinds of features: assembly features, API call features, and string features.
After the features are extracted, they will be used to craft the prompt, which will be then sent to the LLM, querying whether the basic block is part of the implementation of TADA and suitable for a breakpoint.

\subsection{Sample unpacking and BB extraction} \label{sec:method:unpacking}
Since the majority of the malware executable samples are Windows Portable Executable (PE) files, we mainly focus on PE in this work.
However, it is very common to see a PE malware packed~\cite{muralidharan2022file}.
Without unpacking, one can never construct the CFG reflecting real malware logic.
Therefore, the first step of our workflow is to unpack the malware samples and construct the CFG.

The topic of unpacking PE files~\cite{martignoni2007omniunpack, royal2006polyunpack, coogan2009automatic} and constructing CFG has been well studied and many techniques are available.
For the purpose of self-containing, here we only briefly introduce the techniques adopted during our experiments when handling real-world malware samples.
Given a malware sample, we first check if it is a PE file by checking the magic bytes and parsing the header.
If it is a valid PE file, we will then use \texttt{Detect-It-Easy} tool to find whether the file is packed.
In cases where no known packer is found, we will then check the Import Address Table (IAT).
If there are fewer than 5 imported libraries or 15 imported functions, we will also view this sample as a packed malware sample.

As for unpacking, if the malware is packed by UPX, we will just go ahead and unpack it using the UPX utility.
For other packers, we will do dynamic unpacking.
We first will try automatic unpacking using \texttt{PE-sieve} tool.
Essentially, after the malware is loaded in the memory, the tool will continuously scan the memory to check whether the executable code is inconsistent with the copy on the disk.
If any suspicious changes in executable segments are detected, the tool will dump the executable, which is very likely to be the unpacked malware sample.
Unfortunately, in some cases, this automatic unpacking will fail due to corrupted IAT in the dump, and here our last try is to do manual unpack using \texttt{x64dbg}, and fix the imports using \texttt{Scylla} plugin.

To this point, if the original IAT can be parsed, constructing CFG and dumping the BBs are fairly straightforward.
Existing tools include but are not limited to \texttt{IDA}, \texttt{Angr}, etc.

\subsection{Feature Construction: Assembly Feature} \label{sec:method:asm}
Assembly features are features extracted solely from the assembly language, which can be acquired directly from the extracted BBs.
In particular, we are looking for two specific pieces of information: mnemonics of instructions and memory accesses.

In assembly language, mnemonics of instruction are the most semantic-rich component, and the presence of certain mnemonics can be a strong indicator of TADAs.
For example, if a BB contains both \texttt{pushf} and \texttt{popf} instructions, then it is extremely suspicious to detect a debugger.
For another example, if a BB contains hardware querying instructions such as \texttt{cpuid}, then it could be detected if a VM exists.
As for memory accesses, we particularly look for memory accesses through segment registers.
On X86 Windows host, many processes-related information can be accessed through segment registers.
For example, the process environment block (PEB) can be accessed through \texttt{fs} register, which contains various information that can indicate whether a process is being debugged or traced.

Accordingly, for each BB, we iterate through each instruction and check whether the instruction has uncommon, TADA-related mnemonics.
In our experiments, we include in total of 15 uncommon mnemonics, which are shown in Appendix \ref{appendix:augment}.
If any uncommon mnemonics are found, we will report this finding in the prompt, such as:
\begin{displayquote}
    \footnotesize
    \texttt{Uncommon INS: \textlangle mnemonic\textrangle}
\end{displayquote}
We also check whether there is any segment register reference, if so, we will report the finding in the prompt in a similar manner:
\begin{displayquote}
    \footnotesize
    \texttt{Segment Register Access: \textlangle reg\textrangle :\textlangle offset\textrangle}
\end{displayquote}

Assuming LLMs are true oracles, the information provided above should be sufficient for the assembly feature.
Unfortunately, based on our experiment, the result was not satisfying.
From human analysts' perspectives, interpreting and understanding the assembly language features in the context of TADA identification is not difficult, as long as the binary can be disassembled correctly (i.e. no sophisticated anti-static analysis).
However, digesting the assembly features for LLMs can be challenging, since most of the words and phrases in assembly language are abbreviated to an extremely short and less informative form.
Such abbreviation is not an issue for humans, because once human analysts discover some uncommon mnemonics or register accesses, they will be able to refer to other resources such as processor manuals or OS manuals to find out whether they are related to anti-dynamic analysis.
As for LLMs, if we do not provide enough explanation in the prompt, LLMs are essentially seeing this information without any context.
During our experiments, we found that this could cause some of the phrases to be misinterpreted by the model.
For instance, \texttt{STR} instruction x86 is Store Task Register, but when the phrase \texttt{STR} pops up in any computer-related discussion, it is almost certain that the \texttt{STR} means strings.

Therefore, after collecting the reported mnemonics and segment register accesses, we augmented the feature by including an explanatory description of the mnemonics and the register accesses, such as:
\begin{tcolorbox}[colback=blue!5!white, colframe=blue!75!black, sharp corners]
\footnotesize

Uncommon INS: \textlangle mnemonic\textrangle\ (explanations) \\
Segment Register Access: \textlangle reg\textrangle :\textlangle offset\textrangle\ (explanations)

\end{tcolorbox}
The details of the assembly feature augmentation are included in Appendix \ref{appendix:augment}.

\subsection{Feature Construction: String Feature} \label{sec:method:string}
\begin{figure}[t]
    \centering
    \includegraphics[width=0.98\linewidth]{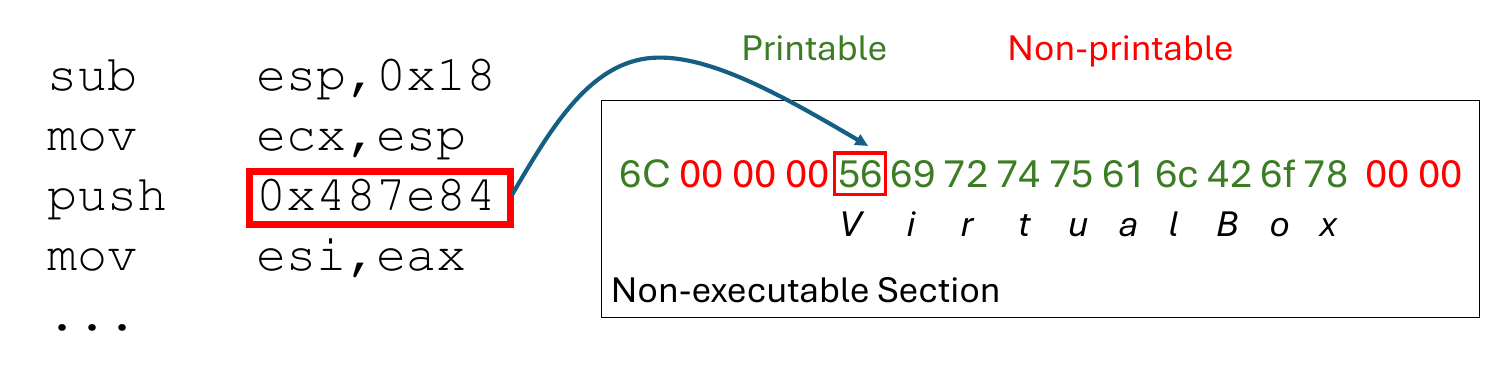}
    \caption{Example of String Reference}
    \label{fig:feature_string}
\end{figure}
One key motivation for using LLM is that the string can carry many clues regarding whether a BB is TADA-related.
For example, as shown in \autoref{fig:feature_string}, if a BB refers to a string such as \textit{VirtualBox}, then obviously the BB is very suspicious of involving TADA.
Accordingly, we include string features, such as:
\begin{tcolorbox}[colback=blue!5!white,colframe=blue!75!black]
{ \footnotesize
    String Reference: "STRING"
}
\end{tcolorbox}

\begin{figure}[t]
    \centering
    \includegraphics[width=0.88\linewidth]{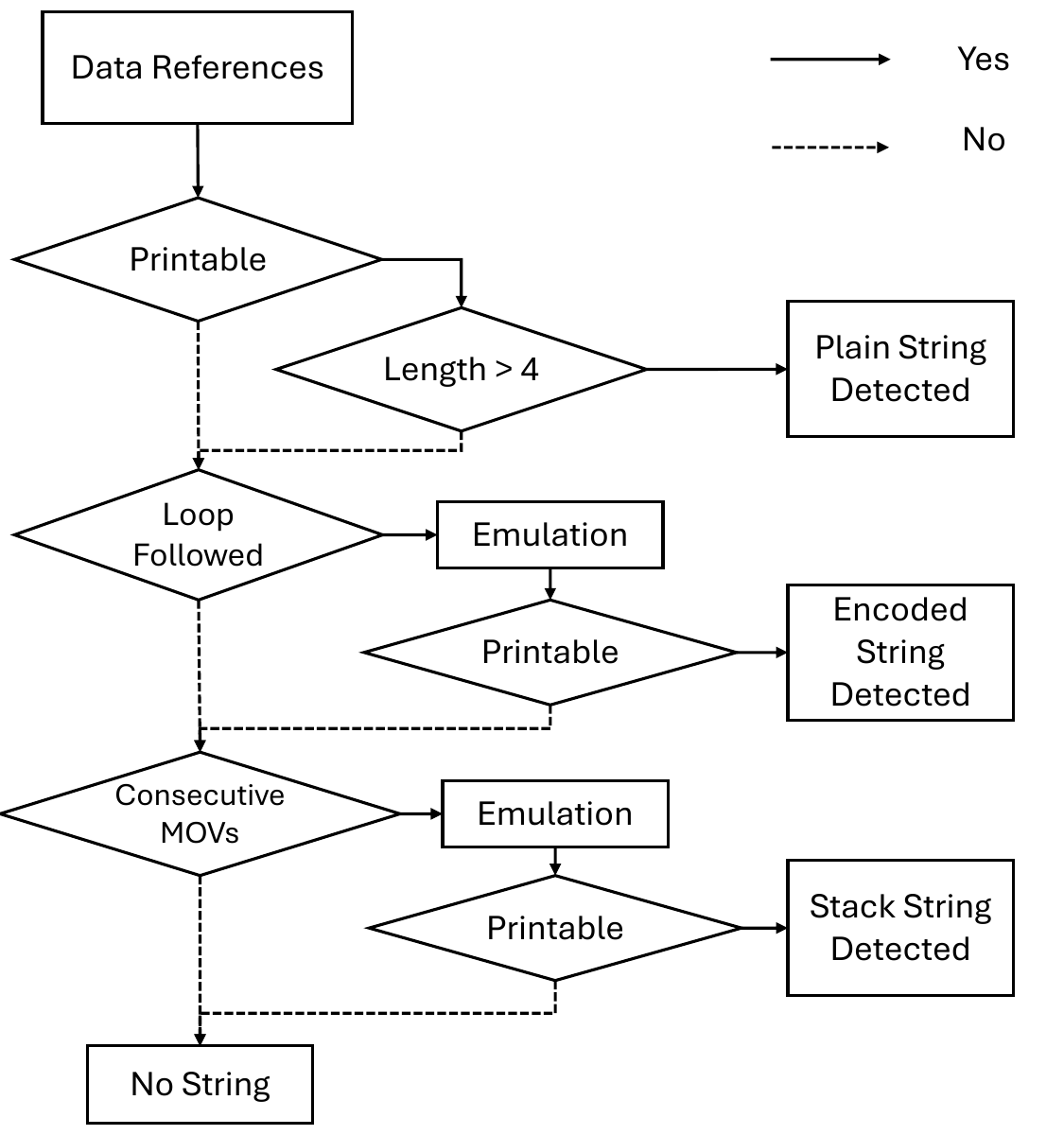}
    \caption{String Deobfuscation}
    \label{fig:string_deobfs}
\end{figure}

However, in the context of malware sample analysis, extracting strings through static analysis could be challenging, because most of the malware, especially those sophisticated ones, will try to hide and obfuscate the strings.
The two most popular strategies to obfuscate strings are string encoding and stack strings, as well as use both strategies simultaneously.
Accordingly, we consider the following 4 situations: 1) string is not obfuscated; 2) string is encoded; 3) string is constructed on the stack and 4) string is encoded and is constructed on the stack.

\autoref{fig:string_deobfs} shows the overall logic of our string deobfuscation process.
First, we check whether the data reference is a plain string: a string saved in plain text format with no obfuscation.
If not, we will emulate the function whenever we suspect the data reference is referring to data that could be a string.
We have two major criteria to determine whether the emulation will be performed: 1) whether a loop, especially a single BB loop is presented; and 2) whether there are consecutive \texttt{mov} instructions (more than 5) that are moving data from data sections to the stack.
The intuitions behind the two criteria are simple: for encoded strings, a loop is extremely likely to be presented, if not inevitable, as decoding data almost always requires loops; for stack and stack encoded strings, consecutive \texttt{mov} instructions must present to set up the stack.
If any of the criteria is met, we will then do the emulation to find the strings.
After a string is found, it will first be used to construct a string feature as shown above, and then be recorded for later use.

\subsection{Feature Construction: API Call Feature} \label{sec:method:api}
As the name suggested, the API call feature of a BB reflects all the API calls in the BB.
Needless to mention, APIs called during malware execution are one of the most important information to understand the behavior of the sample.
For our purposes, API calls are indeed critical as well, as numerous TADAs can be implemented through APIs.
For example, the most famous single API that can be used to do anti-debugger is \texttt{IsDebuggerPresent}, which will return the value of the \texttt{BeingDebugged} flag in PEB~\footnote{\url{https://learn.microsoft.com/en-us/windows/win32/api/winternl/ns-winternl-peb}}.
For another example, in many anti-sandbox techniques, malware authors will conduct detection based on computer and username, which can be very difficult to acquire without using the Windows APIs such as \texttt{GetComputerName}.
Therefore, APIs is extremely important for detecting TADA.

\begin{figure}[t]
    \centering
    \includegraphics[width=0.9\linewidth]{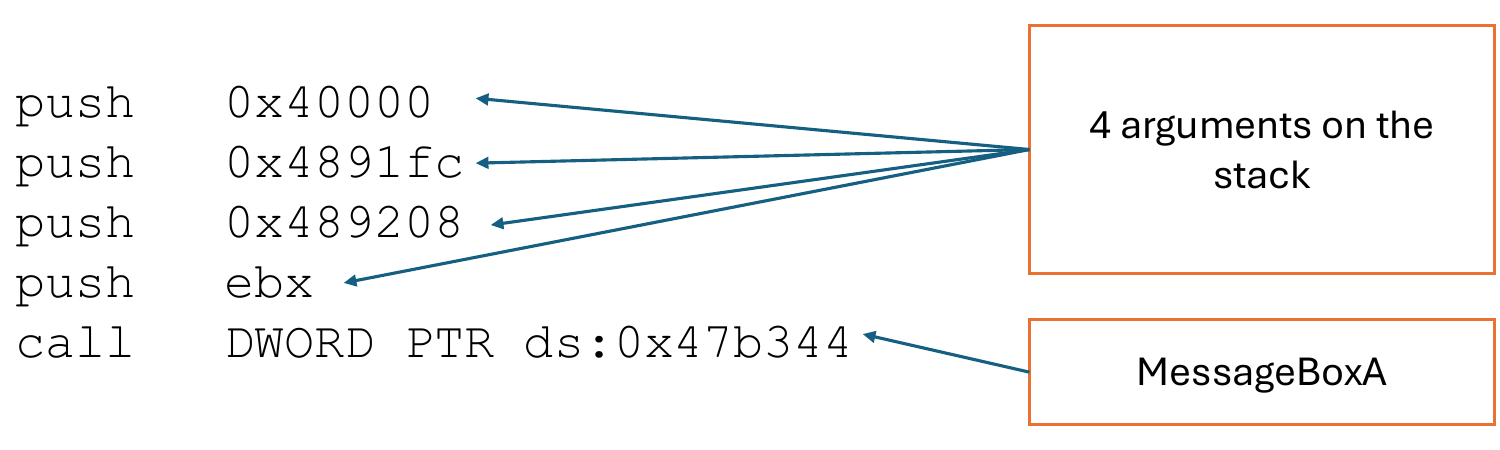}
    \caption{Example of API Call}
    \label{fig:feature_api}
\end{figure}
In our API call feature, we not only include the API name, but also the arguments passed to the API.
Shown in \autoref{fig:feature_api} is an example of API call in a BB calling \texttt{MessageBoxA}.
Since the call instruction refers \texttt{DS} register, after checking the IAT, we will be able to find out that this call is calling \texttt{MessageBoxA}.
Subsequently, based on the documentation of the \texttt{MessageBoxA}, we know that the 4 pushed data are the 4 arguments of the call.

\noindent \textbf{API Name}
We first resolve the name of the API, which is very straightforward in cases of direct calls.
In particular, for each \texttt{call} instruction, we check if the callee is referring to the import table.
From the IAT, we will be able to resolve the API that is being called easily.

Indirect calls are extremely challenging, since our feature extraction only uses static analysis.
However, there are still simple cases where calls may be resolved, as long as the register value at the call site can be inferred.
To do so, if we find an indirect call instruction in a BB, we will construct the data flow graph (DFG) for the function containing the BB.
Then we trace the data flow backward from the indirect call site to see if a live variable with a concrete value can be found (e.g. value from static data sections).
If so, we will check if the value is a valid address for an API, in a similar way as described earlier for direct calls.

\noindent \textbf{Arguments}
Based on Windows \texttt{\_\_stdcall} calling convention and API documentation, it is not difficult to resolve registers of arguments, but what is more important is the values of the arguments.
Similar to what we have done in resolving the API name, we could find the immediate values if they are available.
In addition, we also want to resolve the static and stack strings if the immediate values of arguments are pointers pointing to strings, because, similar to humans, strings could provide a lot of information for LLMs, which is also the key motivation of this work.
If an argument's value appears to be a valid address, we will check if any string is presented at the address.

The constructed API call feature will be in a similar form of calling a function in programming languages, such as:
\begin{tcolorbox}[colback=blue!5!white,colframe=blue!75!black]
{ \footnotesize
    API\_Name(arg1, arg2)
}
\end{tcolorbox}

\subsection{Leveraging LLM}
\noindent \textbf{Input.}
Our ultimate goal is to help the reverse engineers identify the location of the TADA implementations, so that breakpoints can be properly set.
As described in \autoref{sec:method:overview}, we would want to ensure the prompt consists mainly of natural language, so that the LLM can be used without fine-tuning and few-shot learning.
Therefore, our prompt design starts with an initial prompt in plain English of the instructions and the specifications of the TADAs, followed by the features extracted in the form of natural language.

\noindent \textbf{Output.}
At first glance, it may seem that the most intuitive model of this problem is binary classification: having the LLMs to output a binary verdict of whether a BB is TADA related or not.
However, as described in \autoref{sec:problem:llm}, identifying implementation of TADA is a very challenging task and is very prone to false positive, so that the best scenario for a human analyst is to have quantified results which can give them a sense of which BB should be prioritized for investigation.
Instead of prompting the LLM to output a binary result, we prompt for a rating.
The LLM will essentially output a number from 0 to 10, reflecting how related a BB may be related to the implementation of TADA.

The template of the prompt is shown below:
\begin{tcolorbox}[colback=blue!5!white,colframe=blue!75!black]
{ \footnotesize
I want you to help me identify whether a basic block in a binary program is related to anti-dynamic analysis techniques, such as detecting a debugger, sandbox and/or VM.
I will provide some static analysis results of the basic block, including 1) Called APIs (API), 2) Static Strings referred, 3) Uncommon instructions (INS), and 4) Segment Register Reference (SegReg)
Rate from 0 to 10, how likely the code is related to anti-analysis.
\\
\\
I will use your answer to decide whether to put a breakpoint at the basic block, so try to avoid false negatives, and DO NOT consider anti-static analysis techniques.
Please only give the rating number, no explanation
\begin{itemize}
    \item Feature 1
    \item Feature 2
    \item ...
\end{itemize}
}
\end{tcolorbox}

We set a threshold at 7 for positive results, that is, BBs that receive a rating of 7 are considered to be related to TADA implementation.

%% file: sections/evaluation.tex
\section{Evaluation} \label{sec:evaluation}
As described in Section \ref{sec:problem}, the major goal of our work is to ease the human labor involved in identifying the locations of TADA in a malware executable binary.
Accordingly, we want to focus on three research questions: 1) Can our method identify well-known TADAs? 2) Can our method detect TADA in real-world samples? and 3) How much time our method may save for the human analysts?
The first two questions are essentially evaluating the effectiveness of our method: whether our method could identify the TADA after all; the third questions is evaluating the amount of human labor may be saved quantitatively.

Our implementation is based on the workflow introduced in Section \ref{sec:method}, which will be elaborated in the next subsection.
The output of our workflow is binary for each of the BB: whether a breakpoint should be placed at the beginning of the BB.
Every BB that receive an LLM rating of 7 or above is considered as positive BB and a breakpoint shall be placed.

\subsection{Implementation and Malware Samples}
In our experiment, most of our implementation is written in Python.
In terms of the libraries and tools, we used \texttt{IDAPython} for disassembling, CFG construction, and IAT parsing; \texttt{Miasm} library for DFG construction and data flow analysis; \texttt{FLOSS} library for string deobfuscation emulation.
As for LLM, all our experiment uses \texttt{GPT4-Turbo} from OpenAI.
The input parameters are configured to make the model output to be deterministic.

Regarding the malware samples involved in our experiments, they are all collected from VirusTotal using VirusTotal Intelligence Query.

\subsection{Can our method identify well-known TADAs?}
Despite the diversity of TADA mentioned in \autoref{sec:problem:llm}, there still exists many popular and well-known TADAs, which are commonly adopted by malware authors.
Thus, we first need to ensure that our method can identify the BBs implementing these popular TADAs before we explore those interesting tricky ones.
We built a set of programs, each of which includes an implementation of one well-known TADA.
Instead of implementing everything from scratch, we adopted publicly available implementations of these TADAs from a popular GitHub repository of malware anti-analysis techniques, \texttt{al-khaser}.

In total we collected 164 implementations of TADA from \texttt{al-khaser}, and created 164 programs accordingly.
According to the categorizations introduced in \autoref{sec:background:category}, 164 programs can be categorized by either tactics or implementations of the corresponding TADAs, as shown in \autoref{tab:n_sample_tada_category}.
\begin{table}[ht]
    \centering
    \begin{tabular}{ll}
    \toprule
    Tactic              & Count \\
    \midrule
    Debugger Evasion        & 35           \\
    Sandbox Evasion       & 47           \\
    VM Evasion            & 46           \\
    Analysis Tool Evasion & 36           \\
    \midrule
    Total                   & 164       \\
    \bottomrule \\
    \end{tabular}
    \quad
    \begin{tabular}{ll}
    \toprule
    Implementation              & Count \\
    \midrule
    Assembly        & 19           \\
    Direct API       & 13           \\
    Indirect API            & 132           \\
    \midrule
    Total                   & 164       \\
    \bottomrule \\
    \end{tabular}
    \caption{Number of Created Programs Per Category For Different Categorization}
    \label{tab:n_sample_tada_category}
\end{table}
Since all the TADAs are self-implemented, the compiled programs are not packed.
All 164 programs are analyzed using the workflow described in \autoref{sec:method}, except for the malware unpacking components.
We consider an implementation of TADA to be detected if and only if \textit{at least one BB that is a part of the implementation of the TADA is reported by our tool as positive (e.g., LLM rating greater than 7)}.
Therefore, in this experiment, the maximum number of positive BBs can be detected are far more than 164, in cases where the implementation of a TADA involves many BBs.
However, for the sake of answering the research question, we will discuss our result in this experiment in terms of the number of the TADA implementation detected, not the number of BBs.


\begin{figure}[ht]
    \centering
    \includegraphics[width=0.9\linewidth]{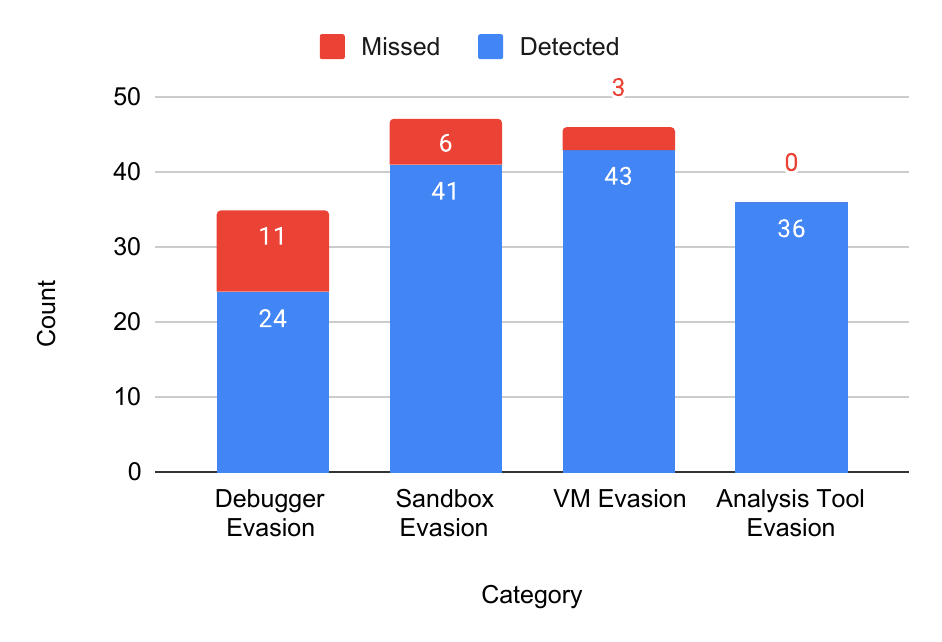}
    \caption{Detection Rates Per Tactic}
    \label{fig:repo:purposes}
\end{figure}

\begin{figure}[ht]
    \centering
    \includegraphics[width=0.9\linewidth]{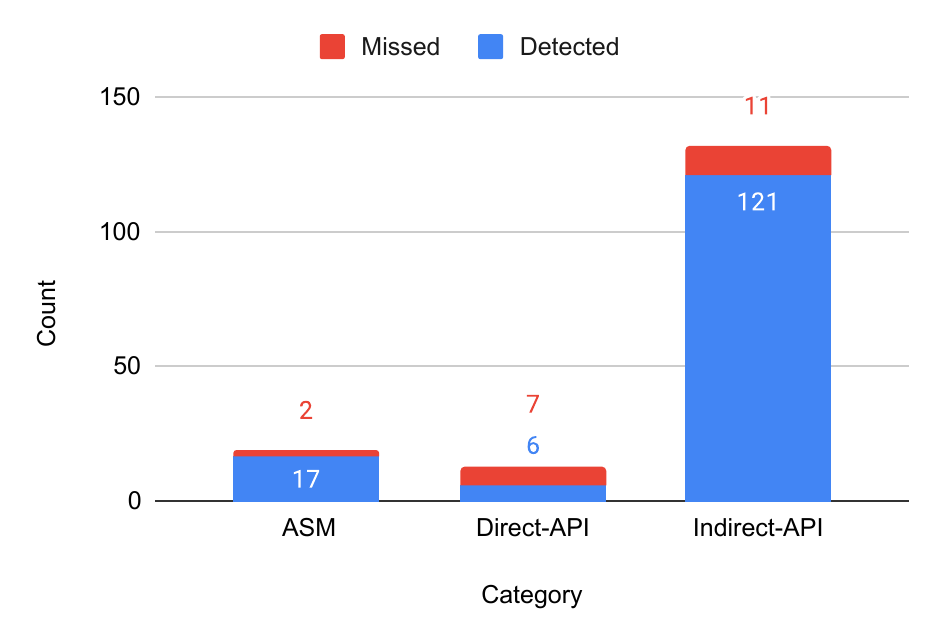}
    \caption{Detection Rates Per Implementation}
    \label{fig:repo:impl}
\end{figure}

In summary, out of all the 164 TADA implementations, we have successfully detected at least one BB in 144 of them, yielding a detection rate of 87.80\%.
To conduct the root causes analysis for the false negatives, we start with investigating the detection rate per category.

Shown in \autoref{fig:repo:purposes} is the detection rates per-tatic category. It is noteworthy that out of 35 debugger evasion TADAs, only 24 of them are detected and 11 of them are false negatives, rendering a detection rate of merely 68.57\%.
This number is particularly low compared to the average detection rate mentioned earlier.
After we studied the false negatives, we found one of the major reasons is that the indirect calls are not accurately resolved by our tool, which can be clearly reflected in \autoref{fig:repo:impl}.
The second reason is that some API calls are not directly related to the anti-dynamic-analysis, causing the LLM to give conservative ratings for those BBs calling those APIs.
Back to the causes of the particularly low detection rate of the debugger evasion category, the reason to be blamed is that most of debugger evasion TADAs can be implemented purely using single or few APIs.

\autoref{fig:repo:impl} shows the detection rates per-implementation category.
Percentage wise, our tool did badly on direct-API based ones.
The root cause mentioned earlier, that our tool did not resolve the indirect calls.
As for the 11 false negatives in indirect-API and 2 false negatives in assembly, it is due to the API called or the instruction used is not directly related to the anti-dynamic-analysis.

Lastly, we reflect to one of our key motivation of using LLM.
As elaborated in \autoref{sec:problem:llm}, we observed that the majority of TADAs, especially tricky ones, usually involve strings.
\begin{table}[ht]
    \centering
    \begin{tabular}{llll}
    \toprule
         & Total Number & Detected & Detection Rate \\
    \midrule
    Involving String            & 111          & 110      & 99.10\%        \\
    No String             & 53           & 34       & 64.15\%       \\
    \bottomrule \\
    \end{tabular}
    \caption{Detection Rates on TADAs Using Strings}
    \label{tab:result_string}
\end{table}
As shown in Table \ref{tab:result_string}, about 2/3 of the TADAs involve strings, and our tool only failed to detect only 1 of them.

\subsection{Can our method detect TADA in real-world samples?}
It is also very important that our method can detect those TADAs in the real-world malware samples, especially those interesting TADAs. 
Here, \textit{interesting} TADAs refer to those TADAs bringing "surprise", which will inspect unpopular and unconventional hardware and/or system artifacts.
In this section, we study 4 famous real-world malware families that are known to have TADAs.
Some of the techniques are introduced and documented in the analysis report due to their creativity and uniqueness.
Our tool not only successfully detected the documented TADA in the sample, but also detected some techniques that are not documented in detail.
For those who is interested in reproducing the experiment, the hashes are included in Appendix \ref{appendix:hashes}.

\subsubsection{Anti-emulation in Raspberry Robin}
We have done analysis on samples of Raspberry Robin (RR), which is a very active malware family.
In late 2023, analysts found some RR samples were adopting an interesting method to detect Windows Defender Emulation engines~\cite{raspberry_robin}.

Running our tool against the sample with anti-emulation techniques, we identify a BB that is not so obvious to be related to TADA. This BB mainly consists of the following features:
\begin{tcolorbox}[colback=blue!5!white,colframe=blue!75!black,title=Features for Basic Block 0x1000137C]
{ \footnotesize
\begin{itemize}
    \item String Reference: "KeRNel32.DLl"
    \item Called API: \\GetModuleHandleA("KeRNel32.DLl")
    \item String Reference: "MpReportEventEx"
    \item Called API: GetProcAddress(ModuleHandleA, "MpReportEventEx")
\end{itemize}
}
\end{tcolorbox}
After researching, we found a blog~\cite{raspberry_robin} which indicate the basic block is related to detecting Windows Defender Emulation engine.

\subsubsection{Disk Serial Number Check in Zebrocy}
An well-known malware family called Zebrocy adopted hardware and environment checking in order to detect the sandbox environment.
According to an online analysis blog~\cite{vast_ocean}, the malware will not exhibit malicious behavior if one of commonly used disk serial numbers is found.

Running our tool against the sample shown in the blog, we are able to detect the first BB of the implementation of this TADA technique. The BB mainly contains following features:
\begin{tcolorbox}[colback=blue!5!white,colframe=blue!75!black,title=Features for Basic Block 0x47ABA0]
{ \footnotesize
\begin{itemize}
    \item String Reference: "c:\textbackslash"
    \item Called API: GetVolumeInformationA("c:\textbackslash\textbackslash", 0, 0, \&VolumeSerialNumber, \&MaximumComponentLength, \&MaximumComponentLength, 0, 0)
\end{itemize}
}
\end{tcolorbox}

We confirmed our findings based on the blog post. It is shown that this is the BB querying the disk serial number, followed by a series of comparisons with known disk serial numbers used by sandboxes.

\subsubsection{Comprehensive Hardware Checks in Trickbot}
Trickbot is a famous malware family that is very popular around the year of 2020. It is a trojan mainly spread by email, and has been thoroughly studied by the authors of~\cite{vast_ocean}.
The most interesting fact of Trickbot in terms of its anti-dynamic-analysis behavior is that this family will stare at the hardware information, carefully and extensively.

Our tool study the sample with hash listed in~\cite{vast_ocean}, and found several BBs that are directly related to TADA. For examples, we found a BB contains code querying WMI:
\begin{tcolorbox}[colback=blue!5!white,colframe=blue!75!black,title=Features for Basic Block 0x404E25]
{ \footnotesize
\begin{itemize}
\item String Reference: \\"SELECT * FROM Win32\_BIOS"
\item String Reference: "W"
\end{itemize}
}
\end{tcolorbox}

In addition, this sample also tried to detect virtual environments. One example is shown in the following BB:
\begin{tcolorbox}[colback=blue!5!white,colframe=blue!75!black,title=Features for Basic Block 0x404F6A]
{ \footnotesize
\begin{itemize}
\item String Reference: "VMWare"
\item Called API: \\StrStrIW(pvarg.bstrVal, "VMWare")
\end{itemize}
}
\end{tcolorbox}

Bios features are also checked:
\begin{tcolorbox}[colback=blue!5!white,colframe=blue!75!black,title=Features for Basic Block 0x404FBE]
{ \footnotesize
\begin{itemize}
\item String Reference: "A M I"
\item Called API: \\StrStrIW(pvarg.bstrVal, "A M I") )
\end{itemize}
}
\end{tcolorbox}

\subsubsection{Analysis Tool Searching in Mustung Panda}
During our real-world sample analysis, we have also seen some old-school yet still very effective methods against dynamic analysis.
One example is Mustung Panda family, which adopted multiple TADA.
Firstly, Mustung Panda use \texttt{CPUID} instruction to query processor information. The BB which performs this technique will have features as follows:
\begin{tcolorbox}[colback=blue!5!white,colframe=blue!75!black,title=Features for Basic Block 0x41CD4B]
{ \footnotesize
\begin{itemize}
\item Uncommon INS: cpuid
\end{itemize}
}
\end{tcolorbox}

Besides, Mustung panda samples not only check for hardware such as processor, but also search for reverse engineering tools. One BB implementing such technique has features as follows:
\begin{tcolorbox}[colback=blue!5!white,colframe=blue!75!black,title=Features for Basic Block 0x401510]
{ \footnotesize
\begin{itemize}
\item String Reference: "cheatengine-i386.exe"
\item String Reference: "WPE PRO.exe"
\item String Reference: "Fiddler.exe"
\item ...
\end{itemize}
}
\end{tcolorbox}
Both BB are among the detected BBs by our tool with high confidence.


\subsection{How much time our method may save for the human analysts?} \label{sec:evaluation:scaled}
This is a difficult question to answer directly in quantitative manner.
Therefore, instead we evaluate this question indirectly by investigating the number of BBs that our tool can find out of a program.

In this experiment, we use real-world malware samples collected from Virus Total using the Virus Total intelligence search.
Specifically, we query the Virus Total for samples that are Windows PE, at least 20 vendors give malware/adware verdict and with anti-analysis tags (e.g. attack technique T1622 or T1497).
In addition, samples are from April 2024 to June 2024, so that they are relatively new at the time of writing and evaluating.
To make our analysis statistically significant, we have sampled in total 30 malware/adware samples from Virus Total using the query described earlier.


We first dissemble the samples and compute the total number of BBs of each sample.
On average, the total number of BBs in a sample is 6869.31, and after analyzing all 30 samples using our tool, we found that on average we can find 12.57 BBs positive for TADA implementations.
In these 30 samples, the number of detected BBs is significantly smaller than that of the total number of BBs, whose ratio is about 0.17\%.
We believe this ratio indicates that finding code of TADA is really finding a needle in a haystack, and our tool should be able to save a good amount of time for the human analysts.

%% file: sections/discussion.tex
\section{Discussion}
\subsection{TADA in benign software}
The presence of TADA does not necessarily mean that a sample is malware.
In fact, many benign samples and "gray" samples such as adware will also have advanced TADA.
For example, the binary executables for online games usually implement TADAs to detect not only debugger and cheating software, but also virtual machines and emulators, in order to protect their games against cracking for piracy.
Therefore, although there are academic researches~\cite{aboaoja2023dynamic, kirat2015malgene} focusing on detecting malware partially based on the presence of TADA, this practice is extremely rare in the real world, which is prone to false positives. 

\subsection{Limitations}
\noindent \textbf{Limitation of static analysis.}
Since we adopt static analysis, our method shares common limitations of static analysis.
This include resolving pointer alias (of indirect calls), vulnerable to file patching and packing obfuscations, and accuracy issues. 
Nevertheless, as shown in \autoref{fig:malware_detection_system}, our scope focuses on reverse engineering, so static analysis is an inevitable step.
Additionally, in principle, our methodology does not prevent us to extract features using dynamic analysis, and therefore, this would be the future work.

\noindent \textbf{Limitations of file types.}
While we are focusing on detecting TADAs in the executable, it is also note-worthy that many anti dynamic analysis efforts may be performed in different file types, such as VBA scripts and JavaScripts.
For different file types, it is necessary to re-design the features that may be extracted from the file, which could be a future direction of this field.

\noindent \textbf{Limited LLM prompt engineering.}
In this work, we primarily focus on specifying the definition of the TADA in the prompt, and we did not adopt any advanced prompt engineering methods.
One future work is to use dynamic few-shot learning during the prompt crafting phase, so that an example set of features of TADA related BB can be added in the prompt.
This will enable the LLM to learn from the ground truth of existing known TADA BBs.




%% file: sections/relatedwork.tex
\section{Related Work}

\subsection{TADA detection}
Anti-debug techniques are used for malware to detect debuggers when malware decides if it stops running and prevents being analyzed.
Cobra \cite{vasudevan2006cobra}, focused on self-modifying and self-checking code to mitigate anti-debug techniques.
Roundy and Miller \cite{Roundy2010HybridAA} developed a hybrid analysis technique that combines static and dynamic analysis to overcome anti-debugging and other evasion techniques used by malware.
Sunjun Lee et al. \cite{SunjunLee2024HybridDA} proposed a dynamic binary instrumentation approach to detect and bypass anti-debugging techniques in malware analysis.
ZeVigilante \cite{Alhaidari2022ZeVigilanteDZ} presented a machine learning-based approach to detect the presence of anti-debugging techniques in malware samples.
Xue et al. \cite{Xue2017MaltonTO} introduced Malton, a binary analysis platform designed to handle various anti-analysis techniques, including anti-debugging methods.
Biondi et al. \cite{Copty2001EfficientDI} developed a formal verification approach to prove the effectiveness of anti-debugging techniques and their countermeasures.
\cite{chen2008towards} presented a comprehensive overview of advanced anti-debugging and anti-reverse engineering techniques used by modern malware, including timing-based detection and hardware breakpoint detection.

\subsection{Transparent sandbox environment}
Another interesting approach is to make good use of the virtual environment. \cite{balzarotti2010efficient} compares the malware behavior between running in the emulated environment and referenced host, thus detecting anti-debug techniques.
Disarm\cite{lindorfer2011detecting} extended this approach but made a comparison in four emulated environments since it is challenging to get a robust comparison in a non-deterministic environment. \cite{park2019automatic} tries to find specific anti-debug techniques (e.g., API-based, instruction-based) that utilize Intel Pin for window environments.
In order to solve the environment issue, BareCloud\cite{kirat2014barecloud} runs the malware in a transparent bare metal environment without any in-guest monitoring.

\subsection{LLM for binary reverse engineering}
Binary reverse engineering is challenging since there is a lack of high-level semantics even with the help of LLM. Current LLM-based methods use a pre-train approach to learn semantics based on the binary opcode and operand corpus to help downstream tasks, such as variable name and type inference\cite{banerjee2021variable, gpt3}, function similarity detection \cite{pei2020trex, wang2022jtrans}, and vulnerability detection \cite{liu2023harnessing, ye2024detecting}. 
Trex\cite{pei2020trex} learns the execution semantics of dynamic traces and assembly code to detect similar functions. 
For vulnerability detection, \cite{liu2023harnessing} proposes a taint analysis technique combined with LLMs for binary vulnerability inspection. SLFHunter\cite{ye2024detecting} leverages LLM to analyze sensitive custom DLLFs separately to help existing tools discover Command Injection vulnerabilities.
BINSUM\cite{jin2023binary} introduces a systematic approach for LLM to comprehend binary semantics.

%% file: sections/conclusion.tex
\section{Conclusion}
In this work, we proposed an LLM workflow that can help reduce the time and labor efforts of reverse engineers when analyzing malware samples by pinpointing the locations of TADAs.
We evaluate our method on a popular public repository of TADA implementations, as well as famous real-world malware samples that are studied in online blogs. 
We successfully identify the locations of 87.80\% of known TADA implementations and identify the interesting TADAs from real-world samples of 4 families.

%% file: sections/appendix.tex
\section{Extra Context Used To Augment Assembly Feature} \label{appendix:augment}
\subsection{Mnemonics}
\begin{itemize}
    \item \texttt{pushf}: Can be used to read/write EFLAGS register
    \item \texttt{pushfd}: Can be used to read/write EFLAGS register
    \item \texttt{popf}: Can be used to read/write EFLAGS register
    \item \texttt{popfd}: Can be used to read/write EFLAGS register
    \item \texttt{pushfq}: Can be used to read/write EFLAGS register
    \item \texttt{popfq}: Can be used to read/write EFLAGS register
    \item \texttt{int}: CPU Interrupt
    \item \texttt{icebp}: Tracing technique, Single Step Exception
    \item \texttt{bts}: Set trap flag when number is exactly 8
    \item \texttt{rdtsc}: Read time-stamp counter
    \item \texttt{sidt}: Access Interupt Descriptor Table
    \item \texttt{sldt}: Access Local Descriptor Table
    \item \texttt{sgdt}: Access Global Descriptor Table
    \item \texttt{str}: Store Task Register
    \item \texttt{cpuid}: Processor information
\end{itemize}

\subsection{Segment Register Access in MS Windows for X86}
\begin{itemize}
    \item \texttt{fs:0h}: Current Structured Exception Handling (SEH) frame
    \item \texttt{fs:4h}: Stack Base / Bottom of stack (high address)
    \item \texttt{fs:8h}: Stack Limit / Ceiling of stack (low address)
    \item \texttt{fs:Ch}: SubSystemTib
    \item \texttt{fs:10h}: Fiber data
    \item \texttt{fs:14h}: Arbitrary data slot
    \item \texttt{fs:18h}: Linear address of TEB
    \item \texttt{fs:1Ch}: Environment Pointer
    \item \texttt{fs:20h}: Process ID (in some Windows distributions this field is used as DebugContext)
    \item \texttt{fs:24h}: Current thread ID
    \item \texttt{fs:28h}: Active RPC Handle
    \item \texttt{fs:2Ch}: Linear address of the thread-local storage array
    \item \texttt{fs:30h}: Linear address of Process Environment Block (PEB)
    \item \texttt{fs:34h}: Last error number
    \item \texttt{fs:38h}: Count of owned critical sections
    \item \texttt{fs:3Ch}: Address of CSR Client Thread
    \item \texttt{fs:40h}: Win32 Thread Information
    \item \texttt{fs:44h}: Win32 client information (NT), user32 private data (Wine)
    \item \texttt{fs:C0h}: Pointer to FastSysCall in Wow64
    \item \texttt{fs:C4h}: Current Locale
    \item \texttt{fs:C8h}: FP Software Status Register
    \item \texttt{fs:CCh}: Reserved for OS (NT), kernel32 private data (Wine)
    \item \texttt{fs:1A4h}: Exception code
    \item \texttt{fs:1A8h}: Activation context stack
    \item \texttt{fs:6E8h}: Real Process ID
    \item \texttt{fs:6ECh}: Real Thread ID
\end{itemize}


\section{Sample Hashes} \label{appendix:hashes}
\noindent Raspberry Robin: \texttt{242851abe09cc5075d2ffdb8e5 eba2f7dcf22712625ec02744eecb52acd6b1bf}

\noindent Zebrocy:
\texttt{091ffdfef9722804f33a2b1d0fe765d2c2 b0c52ada6d8834fdf72d8cb67acc4b}

\noindent Trickbot: \texttt{3bf0f489250eaaa99100af4fd9cce3a 23acf2b633c25f4571fb8078d4cb7c64d}

\noindent Mustung Panda: \texttt{8f9581a80cd18e2bbe33ba0fe29b 778c8125e30f10a81071019dcd46bc3a2d34}